# EMPIRICAL MODE DECOMPOSITION AND NORMAL SHRINK TRESHOLDING FOR SPEECH DENOISING

Mina Kemiha[1]

[1]Department of electronic, Jijel University, Algeria

kemihamina@yahoo.fr

## ABSTRACT

*In this paper a signal denoising scheme based on Empirical mode decomposition (EMD) is presented. The denoising method is a fully data driven approach. Noisy signal is decomposed adaptively into intrinsic oscillatory components called Intrinsic mode functions (IMFs) using a decomposition algorithm called sifting process. The basic principle of the method is to decompose a speech signal into segments each frame is categorised as either signal-dominant or noise-dominant then reconstruct the signal with IMFs signal dominant frame previously filtered or thresholded. It is shown, on the basis of intensive simulations that EMD improves the signal to noise ratio and address the problem of signal degradation. The denoising method is applied to real signal with different noise levels and the results compared to Winner and universal threshold of DONOHO and JOHNSTONE [11] with soft and hard tresholding. The effect of level noise value on the performances of the proposed denoising is analysed. The study is limited to signals corrupted by additive white Gaussian random noise.*

## 1. INTRODUCTION

Recently, a new temporal signal decomposition method, called Empirical Mode Decomposition (EMD), has been introduced by Huang et al. [1] for analyzing data from nonstationary and nonlinear processes. The major advantage of the EMD is that the basic functions are derived from the signal itself. Hence, the analysis is adaptive in contrast to traditional methods such as wavelets where the basic functions are fixed. The EMD has received more attention in terms of applications [2]-[3], interpretation [4]-[5], and improvement [6]-[7]. The major advantage of the EMD is that the basic functions are derived from the signal itself. The EMD is also used in speech denoising [8]. In fact, speech signal noise reduction is a well-known problem in signal processing. Particularly, linear methods such as the Wiener filtering [9] are largely used, because linear filters are easy to implement and to design. However, these methods are not effective when the noise estimation is not possible or when the noise is colored. To overcome these difficulties, nonlinear methods have been proposed and especially those based on Wavelet thresholding [10]-[11]. A limit of the wavelet approach is that the basic functions are fixed, and thus do not necessarily match all real signals. The EMD decomposes a given signal into a series of IMFs through an iterative process called sifting. The EMD can be seen as a type of wavelet decomposition, each IMF replaces the signals detail, at a certain scale or frequency band [4]. The presence of noise in the signal of interest will result in contamination of each of the modes by a greater or lesser fraction of the noise. [8], the idea of denoising is pre-filtering or thresholding (as defined in wavelet denoising) each IMF separately and then completely reconstructs the signal with all IMFs previously pre-treated. The method is seen as a denoising technique that preserves the important contributions of all IMFs.

Soft and hard thresholding are a powerful technique used for removing the noise components by subtracting a constant value from the coefficients of the noisy signal obtained by the analysing transformation. However, such type of direct subtraction results in a degradation of the speech components, another notable work proposes an adaptive assessment threshold denoising method Donoho et al [11] developed a nonlinear shrinkage denoising method for statistical applications. The shrinkage methods rely on the basic idea that the energy of a signal (with some smoothness) will often be concentrated in a few coefficients in signal while the energy of noise is spread among all coefficients.

In this paper, we combine EMD with adaptive thresholding Normal Shrink for this; we exploit the characteristics of the empirical modes from the EMD to study a new approach denoising signals. It is not an easy task to identify and remove these noise components without degrading the speech signal. Due to the frequency characteristics of IMFs, EMD makes it possible to remove these remaining noise components effectively.

The remainder of the paper is organized as follows. Empirical mode decomposition algorithm is introduced in Section 2. denoising principle is presented in Section 3. EMD denoising in Section 4. Results based real speech signals are presented in Section 5. Finally, conclusions are given in Section 6.

## 2. EMPIRICAL MODE DECOMPOSITION

The empirical mode decomposition has been proposed by Huang et al. as a new signal decomposition method for nonlinear and/or nonstationary signals [8]. The EMD decomposes a given signal into a collection of oscillatory modes, called intrinsic mode functions (IMFs), which represent fast to slow oscillations in the signal. Each IMF can be viewed as a sub-band of the signal. Therefore, the EMD can be viewed as sub-band signal decomposition. Conventional signal analysis tools, such as Fourier or wavelet-based methods, require some predefined basis functions to represent a signal. The EMD relies on a fully data-driven mechanism that does not require any a priori known basis. The algorithm operates through the following steps:

1. Initialize the algorithm: $j=1$, initialize residue $r_0(t)=x(t)$ and fix the threshold $\delta$

2. Extract local maxima and minima of $r_{j-1}(t)$

3. Compute the upper envelope $U_j(t)$ and lower envelope $L_j(t)$ by cubic spline interpolation of local maxima and minima, respectively

4. Compute the mean envelope $m_j(t) = \dfrac{(U_j(t)+L_j(t))}{2}$

5. Compute the $jth$ component $h_j(t) = r_{j-1}(t) - m_j(t)$

6. $h_j(t)$ is processed as $r_{j-1}(t)$. Let $h_{j,0}(t) = h_j(t)$ and $m_{j,k}(t)\ k=0,1,......$ be the mean envelope of $h_{j,k}(t)$, then compute $h_{j,k}(t) = h_{j,k-1}(t) - m_{j,k-1}(t)$ until $SD(i) = \sum_{t=0}^{T} \dfrac{|h_{j,i-1}(t)-h_{j,i}(t)|^2}{(h_{j,i-1}(t))^2}$

7. Compute the $jth$ IMF as $IMF_j(t) = h_{j,k}(t)$

8. Update the residue $r_j(t) = r_{j-1}(t) - IMF_j(t)$

9. Increase the sifting index j and repeat steps 2 to 8 until the number of local extrema in $r_j(t)$ is less than 3

The signal reconstruction process x(t), which involves combining the IMFs formed from the EMD and the residual

$$x(t) = \sum_{j=1}^{N} IMF_j(t) + r_N(t)$$

## 3. DENOISING PRINCIPLE

Let $f_j(t)$ be a noiseless *IMF* and $IMF_j$ its noisy version. Consider a deterministic signal $y(t)$ corrupted by an additive Gaussian white random noise, $b_j(t)$ with a noise level $\sigma_j^2(t)$ as follows: $IMF_j(t) = f_j(t) + b_j(t)$ where $j = \{1,...,N\}$ an estimation $\widetilde{f}_j(t)$ of $f_j(t)$ based on the noisy observation $IMF_j(t)$ is given by $\widetilde{f}_j(t) = \Gamma[IMF_j(t), \tau_j]$, where $\Gamma[IMF_j, \tau_j]$ is a pre-processing function, defined by a set of parameters $\tau_j$, applied to signal $IMF_j(t)$. The denoising signal $\widetilde{x}(t)$ is given by: $\widetilde{x}(t) = \sum_{j=1}^{N} \widetilde{f}_j(t) + r_N(t)$. Regarding the preprocessing function $\Gamma$, there are several approaches in this work, we take $\Gamma$ is thresholding.

## 4. EMD DENOISING

The approach is based on the segmentation of the speech signal into segments of 128 samples, each sub-frame is categorised as either signal-dominant or noise-dominant. The classification pertains to the average noise power associated with that particular sub frame is described by the equation (6) then this sub-frame is characterized as a signal dominant sub frame, otherwise a noise dominant one. In case of a signal dominant sub-frame, the coefficients are not thresholded, since it is highly possible to degrade the speech signal, especially for high SNRs. In the case of a noise dominant sub frame, a thresholding is applied [12]

$$\frac{1}{128}\sum_{k=1}^{128}\left|X_k^i\right|^2 \geq \sigma_n^2$$

With $\sigma_n$ the standard deviation of the noise.

### 4.1. NormalShrink tresholding

The proposed method, called normal shrink is computationally more efficient and adaptive because the requirement to assess the threshold parameter depends on the data the threshold is computed from the following equation $T = \beta \dfrac{\hat{\sigma}^2}{\hat{\sigma}_y}$ Where $\hat{\sigma}^2$ present the noise variance estimate by the following equation

$$\hat{\sigma}^2 = \left[\frac{median(|imf|)}{0.6745}\right]$$

with $\hat{\sigma}_y$ the standard deviation of the level in consideration in our case to denoise IMFs, and the scale parameter, it depends on the size and number of the level of decomposition, and described by the following equation (the level wavelet corresponds our work to IMFs $\beta = \sqrt{\log \dfrac{L_k}{j}}$

$L_k$ is the size of the level, which corresponds in our work the width of the segment, is the number of decomposition or the number of IMFs.

### 4.2. Pseudo code denoising

The Empirical Mode Decomposition also provides the decomposition of a signal into different time-scales or IMFs. This means that it is also possible to filter signal components individually instead of the original signal. This suggests that the strategy for signal denoising based on wavelets may also be applied to intrinsic mode functions. Thus, we propose the following procedure for signal filtering:

Entry: noisy signal

Released: reconstructed signal

**Step A**: set the stop criterion of screening and apply EMD to extract the IMF and the residue.

**Step B**: decompose each IMFs band 128 samples, then set test strips or the signal dominant and bands or noise is dominant.

**Step C**: If the signal is dominant, the strip which is kept as is to step E, if not

**Step D**: use a denoising method to clean the segments or noise is dominant

**Step E**: reconstruct the clean signal.

## 5. RESULTS AND DISCUSSION

To illustrate the effectiveness of the denoising method we performed numerical simulations using two databases, presenting the test signals (speech) chosen randomly from from noise.exe database with 8192 samples and a sampling frequency of 8000 Hz, and TIMIT database a sampling frequency of 16000 Hz and 16383 samples, corrupted with white Gaussian noise with variance 1 and mean zero to obtain the noisy signals with different values of signal to noise ratio SNR : 0 dB, 5 dB, 10 dB, 15 dB The SNR is determined to estimate the effectiveness of the method in terms of reducing the noise present in the signals by comparing the input SNR and the SNR at the output, the output SNR is calculated from equation, where $\tilde{x}(t)$ is the denoised signal,

$$SNR_{out} = 10 log_{10} \frac{\sum_{i=1}^{N}((\tilde{x}))^2}{\sum_{i=1}^{W}(x(t)-(\tilde{x}))^2}$$

Figure (1) displays a comparison between a speech signal contaminated by a Gaussian white noise and clean signals obtained by the methods studied in this paper, and Table 1 shows a comparison between the SNR outputs obtained from used denoising method, as shown in Figure (1-c) , the Wiener filter fails to clean the noised speech signal presented in figure (1-b) and allows a low SNR output proving that denoising method derived from the stationary case are ineffective for denoising nonstationaire signals (in our case, the speech signal). As can be Observed, the cleaned speech signal using the universal threshold of DONOHO and JOHNSTONE with soft and hard threshold shown in figure (1-d) and figure (1-e) allows a high SNR output as shown in Table 1, Conversely, this approach has lost too much details Compared to the true model in Figure (1-a), the denoising with the universal threshold of DONOHO and JOHNSTONE contribute a signal degradation. As can be seen, the proposed method allows to have a SNR output greater than the SNR output obtained the universal threshold of DONOHO and JOHNSTONE with soft and hard threshold as shown in Table 1 , and simultaneously, the denoised signal obtained by the proposed method as shown in figure ( 1 - f ) has no degradation of the signal compared to the true clean signal , the empirical mode decomposition combined with thresholding Normal Shrink denoised allows the non-stationary signals without signal degradation. As can be seen, the proposed method allows to have a SNR output greater than the SNR output obtained with the universal threshold of DONOHO and JOHNSTONE with soft and hard threshold as shown in Table 1 , and simultaneously, the denoised signal obtained by the proposed method as shown in figure (1 - f ) has no degradation of the signal compared to the true clean signal , the empirical mode decomposition combined with Normal Shrink threshold allows a great SNR output and cleans the speech signals without degradation.

Table 1. Comparison between true formant frequencies and those obtained via wavelet-based and EMD-based separation methods for a synthetic vowel /a/ by using a frame length of 1024.

| Input SNR (dB) | Output SNR (dB) | | | |
|---|---|---|---|---|
| | Soft treshold | Hard treshold | Winner filter | Proposed method |
| 0 | 7.0562 | 7.5031 | 3.0138 | **8.0128** |
| 5 | 10.5139 | 10.7205 | 7.0597 | **11.3264** |
| 10 | 14.8233 | 15.1697 | 8.9439 | **15.3426** |
| 15 | 18.5250 | 18.8189 | 10.9275 | **19.1137** |

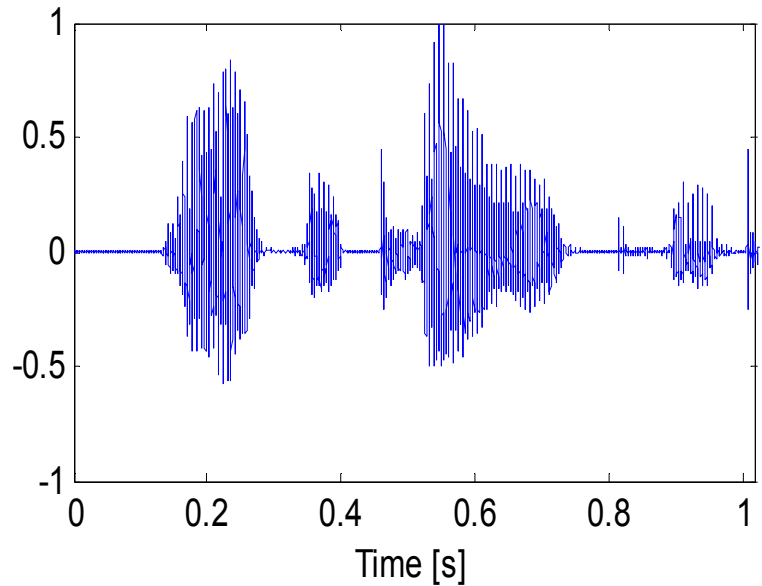

(a)

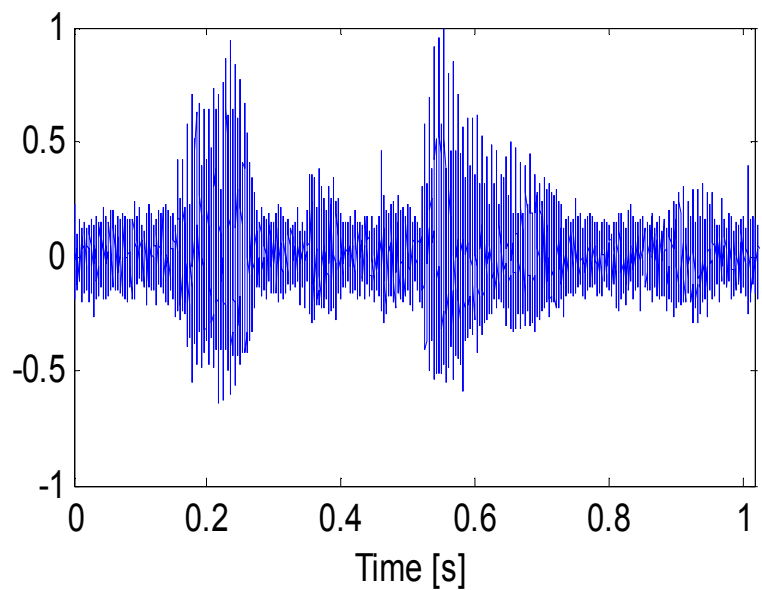

(b)

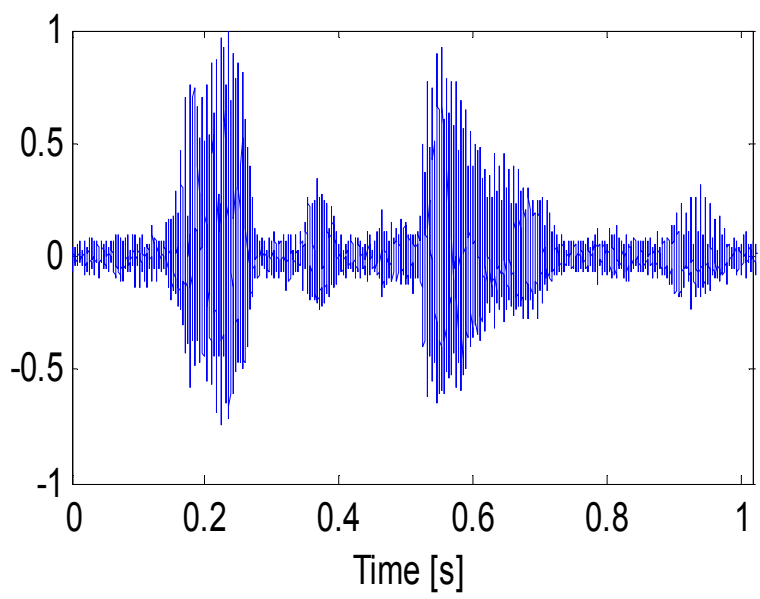

(c)

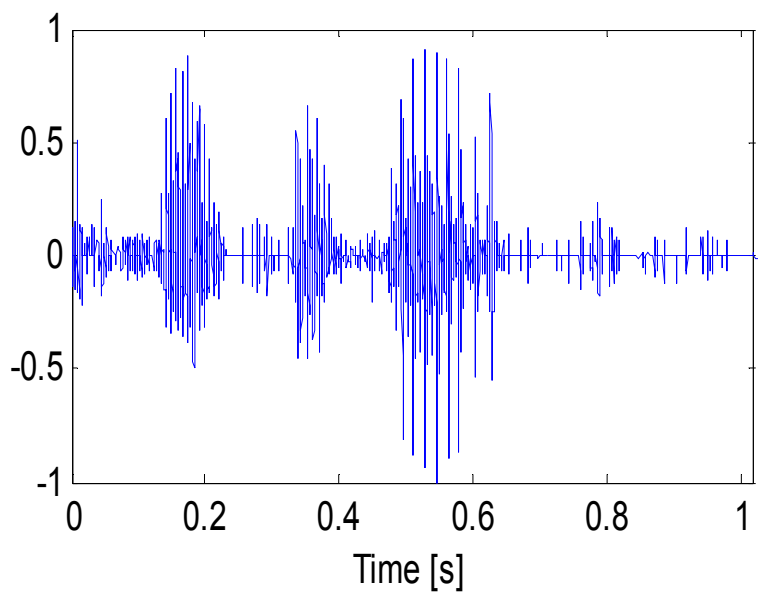

(d)

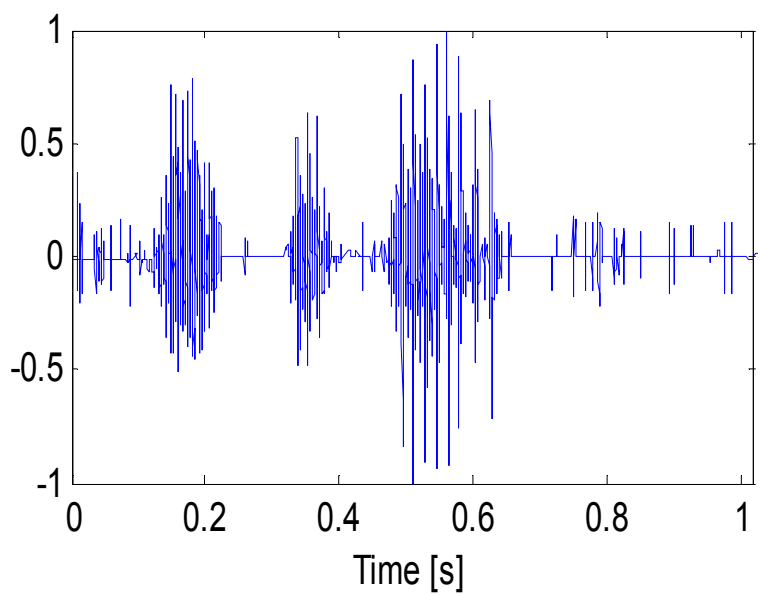

(e)

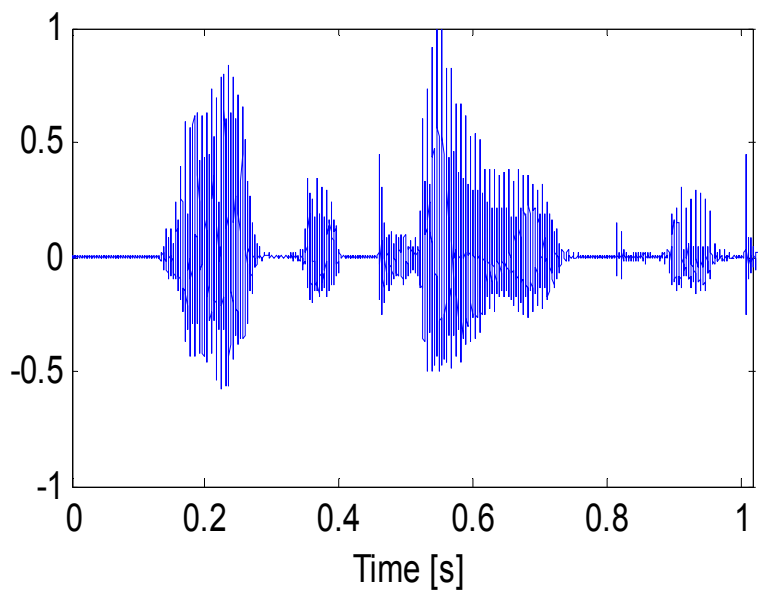

(f)

Figure 1. Waveform of: (a) clean speech, (b) noisy speech at 0dB, (c) denoised speech obtained by Winner filter, (d) denoised speech obtained by Donoho and hard tresholding, (e) denoised speech obtained by Donoho and soft tresholding, (f) denoised speech obtained by proposed method

## 6. CONCLUSIONS

In this paper, a new speech enhancement method to effectively remove the noise components is presented. We have combined two powerful adaptive methods: the EMD and the normal shrink filtering. Obtained results for speech signal contaminated with different noises with different SNR values ranging from 0 dB to 15 dB, showed that the proposed method performs better than the universal threshold of DONOHO and JOHNSTONE with soft and hard threshold and address the problem of signal degradation, the reported results demonstrated that the EMD-Normal Shrink denoising method is effective for noise removal and confirmed that it is a very attractive method to use in general noisy contexts.

## REFERENCES

[1] N.E. Huang and al. "The empirical mode decomposition and Hilbert spectrum for nonlinear and non-stationary time series analysis". Proc. Royal Society, 1998,454(1971):903–995.

[2] F. Salzenstein A.O. Boudraa, J.C. Cexus and L. Guillon. " If estimation using empirical mode decomposition and nonlinear teager energy operator". Proc. IEEE ISCCSP, Hammamet, 2004.,pp 45–48.

[3] A.O. Boudraa S. Benramdane, J.C. Cexus and J.A. Astolfi. "Transient turbulent pressure signal processing using empirical mode decomposition". Proc. Physics in Signal and Image Processing, Mhoulouse, 2007.

[4] P. Flandrin, G. Rilling, and P. Goncalves. "Empirical mode decomposition as a filter bank". IEEE Sig. Proc. Lett., 11(2):112–114, 2004.

[5] Z. Wu and N.E. Huang. " A study of the characteristics of white noise using the empirical mode decomposition method". Proc. Roy. Soc. London A, 2004,460:1597–1611.

[6] B. Weng and K.E. Barner. "Optimal and bidirectional optimal empirical mode decomposition". Proc. IEEE ICASSPToulouse, 2007, 3:1501–1504.

[7] R. Deering and J.F. Kaiser. "The use of a masking signal to improve empirical mode decomposition". Proc. IEEE Philadelphia, 2005.

[8] K. Khaldi, A.O. Boudraa, A. Bouchiki, M. Turki-Hadj Alouane, and E. Samba Diop. "Speech signal noise reduction by EMD". In Proc. IEEE ISCCSP, Malta, March 2008.

[9] J.G. Proakis and D.G. Manolakis. "Digital Signal Processing: Principles, Algorithms, and Applications", volume 1. Prentice-Hall, 3rd edition, 1996.

[10] D.L. Donoho. "De-noising by soft-thresholding". IEEE Trans. Inform. Theory, 1995,41(3):613–627.

[11] D.L. Donoho and I.M. Johnstone. "Ideal spatial adaptation via wavelet shrinkage". Biometrica, 1994,81:425–455.

[12] Erhan Deger, Md. K. Islam Molla, Keikichi Hirose, Nobuaki minematsu and md. kamrul hasan, ''speech enhacemet using soft thresholding with dct-emd based hybrid algorithm", 15th european signal processing conference (eusipco 2007), pp 75-79